\documentstyle[prl,twocolumn,aps]{revtex}
\begin{document}
\draft

\title
{Algebraic Scaling of the Kondo Temperature in a Luttinger Liquid}
\author{Philip Phillips}
\vspace{.05in}

%
\address
{Loomis Laboratory of Physics\\
University of Illinois at Urbana-Champaign\\
1100 W.Green St., Urbana, IL, 61801-3080}

\maketitle

\begin{abstract}
I show here how the algebraic scaling of the Kondo temperature in a
Luttinger liquid can be understood simply
from the condition for the existence of the singlet (or triplet) bound state.
\end{abstract}
\vspace{.1in}

\pacs{PACS numbers:72.10.Fk,71.45.-d, 72.15.Nj}

\narrowtext

In a recent letter Furusaki and Nagaosa (FN)\cite{fn} (and Lee and Toner
(LT)\cite{lt} previously) addressed the problem of the Kondo effect in
a Luttinger liquid.  One of the key results of FN is that the Kondo temperature
scales algebraically as $T_k\propto J^{2\eta}$, with J the Kondo exchange
coupling, $\eta={\pi v_F \over g_2}$, $v_F$ the Fermi velocity and $g_2$ the
coupling constant for forward
electron scattering.  This result is valid strictly in the limit of
weak Coulomb interactions,
${g_2 \over 2\pi v_F }<<1$.  Earlier LT bosonized the extended
Hubbard model and obtained as an exponent $\eta={1 \over
2-K_{\rho}-K_{\sigma}^{-1}}$, the spin-density-wave exponent\cite{lt}.
$K_{\sigma}$ and $K_{\rho}$
are the spin and charge coupling constants for the Luttinger liquid.  Although
this
exponent is valid in the weak-coupling regime, it does not reduce to
the simple form predicted by FN.  FN pointed out that because bosonization does
not preserve the $SU(2)$ symmetry inherent in the Kondo model, exponents
obtained
from this procedure will be in error\cite{fn}. However, the concensus does seem
to be that the Kondo effect in a Luttinger liquid is fundamentally distinct
 from the standard Kondo problem in a Fermi liquid in which the bound singlet
state
forms only when $J<0$ and $T_k$ decays exponentially with the Kondo coupling
as $T_k\propto e^{\left( \frac {-1} {n(\epsilon_F)J} \right)}$.

What I want to show here is that the algebraic scaling of the Kondo temperature
in
a Luttinger liquid obtained by FN has a simple origin and can be predicted
without recourse to
renormalization group analyses. Further, the expression I derive is valid
regardless
of the strength of the electron interactions.  My starting point is the Fermi
liquid problem in which
the zeroth-order Kondo singlet ground state can be written\cite{yosida} as
\begin{equation}
|\Psi\rangle=\sum_{k>k_F}
\alpha_k(a^{\dagger}_{k\uparrow}
\chi_{\downarrow}-
a^{\dagger}_{k\downarrow}\chi_{\uparrow})|F\rangle
\end{equation}
where  $|F\rangle$ is the full Fermi sea
$|F\rangle=\prod_{k<k_F}a^{\dagger}_{k\uparrow}
a^{\dagger}_{k\downarrow}|0\rangle$, $\alpha_k$ is
an expansion coefficient and $\chi_{s}$ is the spin eigenstate of the local
magnetic impurity. The expansion coefficient is easily shown to be given by
$\alpha_k=\langle F|\chi^{\dagger}_{\downarrow}a_{k\uparrow}-
a_{k\downarrow}\chi^{\dagger}_{\uparrow})|\Psi\rangle=\langle
F|b_k|\Psi\rangle$.
The time evolution of $\alpha_k$ can be obtained from the Heisenberg equations
of
motion $i\dot {\alpha_k}=\langle F|[b_k,H]|\Psi\rangle$, where $H$ is the
standard Kondo
Hamiltonian.   If we write $\alpha_k(t)=e^{-iEt}\alpha_k(t=0)$, we find that
the
evolution equation reduces to the stability condition
\begin{equation}
1= -J_o\sum_k^{\prime}{1 \over E-\epsilon_k}
\end{equation}
with $J_o=\frac {J} {N}$. The sum is strictly above the Fermi surface and the
$\epsilon_k$
are the energies of the free conduction states. Eq. (2) is the standard
Cooper-like
 equation, the solution of which yields the
energy (below the continuum) at which the bound singlet Kondo state forms.
Conversion of the sum in (2) to an
integral introduces the single particle density of states.   For the Fermi
liquid
problem, this quantity can be treated as a constant and the right-hand side of
(2)
is easily integrated to yield a logarithm.  The exponential dependence of the
bound
state energy on the Kondo coupling $J_o$ follows immediately.

In the corresponding Luttinger problem, a similar derivation can be followed
because the
bosonization procedure generates a set of non-interacting
psuedo-particles\cite{fdm}.  In this case,
however, the single particle density of states is not a constant.  Rather, it
scales
as a power law and vanishes at the Fermi level.  Below the Kondo temperature,
strong coupling between the
impurity and the electron spin located at the origin
effectively breaks the correlated 1-d electron system into two semi-infinite
Luttinger liquids\cite{fn}.  The Kondo impurity appears as a boundary condition
at one end of the chain.
For the $SU(2)$ symmetric case in which $K_\sigma=1$, the appropriate density
of states for a semi-infinite Luttinger liquid is
$n(\epsilon)\propto |\epsilon|^{{1 \over g}-1} $ in the vicinity of the
Fermi level\cite{kf}. The coupling constant $g$ is the dimensional conductance
in the Luttinger liquid,
$g={1 \over \sqrt{1+{g_2 \over \pi v_F}}}$. For repulsive interactions,
$g_2>0$ and consequently, $g<1$. If we now substitute this form
for the density of states\cite{Anderson} in Eq. (2) and convert the sum to an
integral, we find immediately
that the integral has an algebraic singularity that scales as
$E_b^{-\left( {{1 \over g}-1}\right) }$.
The stability condition for the singlet state now reduces to
\begin{equation}
1=const.J_o E_b^{-\left( {{1 \over g}-1} \right) }
\end{equation}
which implies that
\begin{equation}
E_b\propto J_o^{1 \over  {1 \over g} -1}=k_BT_k.
\end{equation}
Let us now analyze
this expression in the weak-coupling regime.  In this regime $g$ can be
expanded in
a power series in ${g_2 \over \pi v_F}<<1$.
We find that the leading power law dependence,
$T_k\propto J_o^{\left( {2\pi v_F \over g_2} \right)}$,
is precisely the result of FN.  In the strong-coupling regime,
$T_k\propto J_o^{\sqrt {\pi v_F \over g_2 }}$\cite{comment}.  The procedure
of FN is not easily amenable to a strong-coupling analysis.  We find then that
 this simple
procedure based on the analysis of the stability condition for the Kondo bound
state
predicts
algebraic scaling of the Kondo temperature that recovers the weak coupling
result.  It is the power law dependence of the
pseudo-particle density of states for a semi-infinite Luttinger liquid that
leads
to this behaviour.  The physical picture that this
derivation suggests is that the pseudo-particles in the Luttinger liquid
form the screening cloud at the magnetic impurity in direct analogy to the
quasi-particles
in the Fermi liquid problem. Of course this picture changes dramatically if
back-scattering
is included\cite{LE}.  As FN pointed out, in the presence of backscattering,
the two semi-infinite
pieces of the Luttinger liquid become connected and the scaling behaviour
presented
here breaks down.  Away from half-filling, however, back scattering as well as
umklapp processes are not relevant along the Luttinger fixed line.  Hence,
away from half-filling, the analysis presented here should hold true.

\acknowledgments
This work is supported in part by the NSF and the ACS Petroleum Research Fund.
I thank Yi Wan, Nancy Sandler, and A.-H. Castro Neto for insightful remarks.


\begin{thebibliography}{Anderson}
\bibitem{fn}
 A. Furusaki and N. Nagaosa, Phys. Rev. Lett. {\bf 72}, 892 (1994).
\bibitem{lt}
D. H. Lee and J. Toner, Phys. Rev. Lett. {\bf 69}, 3378 (1992).
\bibitem{yosida}
K. Yosida, Phys. Rev. {\bf 147}, 223 (1966).
\bibitem{fdm}
F. D. M. Haldane, J. Phys. C {\bf 14}, 2585 (1981).
\bibitem{kf}
C. L. Kane and M. P. A. Fisher, Phys. Rev. Lett. {\bf 68}, 1220 (1992).
\bibitem{Anderson}
This derivation is in the spirit of the
interlayer tunneling argument used by S. Chakravarty and P. W. Anderson, Phys.
Rev.
Lett. {\bf 72} 3859 (1994).
\bibitem{comment}
The results presented here can be trivially extended to the triplet ground
state,
which also exhibits a Kondo effect in a Luttinger liquid.  In Eq. 1, a plus
sign
should be used and $J\rightarrow -J$.
\bibitem{LE}
A. Luther and V. J. Emery, Phys. Rev. Lett. {\bf 33}, 589 (1974).



\end{thebibliography}
\end{document}